*Letter to the Editor*

# Bipolar jets from the symbiotic star Hen 3-1341 in outburst[⋆]

**Toma Tomov**[1,2], **Ulisse Munari**[1,2], **and Paola Maria Marrese**[1]

[1] Osservatorio Astronomico di Padova, Sede di Asiago, I-36032 Asiago (VI), Italy
[2] CISAS - Center of Studies and Activities for Space, Univ. of Padova "G.Colombo", Italy

**Abstract.** The spectral signature of collimated bipolar jets have been discovered on high resolution spectra of Hen 3-1341 during the 1999 outburst, the first ever recorded for this symbiotic star. The jets show up as emission components at $|\Delta RV_\odot| \sim 820\,\mathrm{km\,s^{-1}}$ on both sides of the main emission lines. They were not present in quiescence. As a general rule for the jets from stellar objects, also those in Hen 3-1341 come together with evidence of mass loss via wind (strong P-Cyg profiles).

## 1. Introduction

Hen 3-1341 ($\equiv$ SS 75) was discovered by Henize (1976) and its classification as a symbiotic star followed the spectroscopic observations by Allen (1978, 1984). The high excitation conditions found by Allen have been later confirmed by Gutiérrez-Moreno et al. (1997), whose optical and ultraviolet spectra show strong emissions in the Balmer continuum, by the N V, [Fe VII], He II lines and the symbiotic band at 6830 Å (attributed by Schmid (1989) to Raman scattering of O VI 1032 Å by neutral hydrogen). Mürset & Schmid (1999) have derived an M4 spectral type for the cool giant in Hen 3-1341, which TiO band dominates the red portion of optical spectra. The UBVRI-JHKL photometric survey of symbiotic stars by Munari et al. (1992) lists $V = 12^{\mathrm{m}}\!\!.94$ for Hen 3-1341, with optical colors typical of symbiotic stars harboring a very hot and luminous white dwarf and infrared colors appropriate to a cool giant without circumstellar dust. The $K = 7^{\mathrm{m}}\!\!.66$ magnitude by Munari et al. is very close to the $K = 7^{\mathrm{m}}\!\!.58$ measured by Allen (1982), suggesting that the cool giant is not variable (like in the majority of known symbiotic stars). These data make Hen 3-1341 to resemble Z And in quiescence, the prototype of symbiotic stars (Kenyon 1986).

The majority of the symbiotic stars undergo outbursts of 2–3 mag amplitude characterized by the disappearance of the high excitation emission lines and the overwhelming of the cool giant spectrum by the hot continuum from the outbursting white dwarf (Ciatti 1982). When we have observed Hen 3-1341 on June 8, 1999 as part of a high resolution spectroscopic survey of symbiotic stars (Munari and Marrese, in preparation), we have found it $\sim 2^{\mathrm{m}}\!\!.5$ brighter than reported for quiescence by Munari et al. (1992), with the high excitation emission lines disappeared and the TiO bands overwhelmed by a featureless hot continuum. We were therefore witnessing the first ever recorded outburst of Hen 3-1341.

However, what makes our observations of Hen 3-1341 in outburst even more interesting is the discovery of high velocity, symmetrically displaced red and blue components in the profile of the emission lines, that we attribute to high velocity bipolar jets. With the same spectrograph we have already discovered similar bipolar jets in the super soft x-ray source RX J0019.8+2156 (Tomov et al. 1998). As for the latter, the jets in Hen 3-1341 are a transient phenomenon: they do not show up in similar high resolution spectral observations by van Winckel et al. (1993) secured in 1989 when Hen 3-1341 was in quiescence.

Jets have already been discovered in the spectra or direct optical/radio images of some symbiotic stars, mainly during outburst phases: R Aqr (Burgarella & Paresce 1992; Dougherty et al. 1995), CH Cyg (Taylor et al. 1986; Solf 1987), MWC 560 (Tomov et al. 1990; Shore et al. 1994), and RS Oph (Taylor et al. 1989).

## 2. Observations

High-resolution spectra ($R \sim 0.30$ Å at H$\beta$) have been obtained with the Echelle spectrograph mounted at the Cassegrain focus of the 1.82 m telescope which is operated by the Padova and Asiago Astronomical Observatories on top of Mt. Ekar, (Asiago, Italy). The detector has been a Thomson THX31156 CCD with 1024×1024 pixels, 19$\mu$m each, and the slit width has been set to 2″. A medium dispersion spectrum ($R \sim 1$ Å) of the H$\alpha$ region has been secured with the B&C+CCD spec-trograph at the 1.22 m telescope of the Astronomy Department, University of Padova, located in Asiago too. The detector has been a Wright Instr. CCD camera with a 512×512 pixels, 23$\mu$m size, UV-coated chip. The slit width has been set to 2″. The reduction and analysis of all the spectra has been performed in a standard fashion under IRAF. A journal of the observations is given in Table 1.

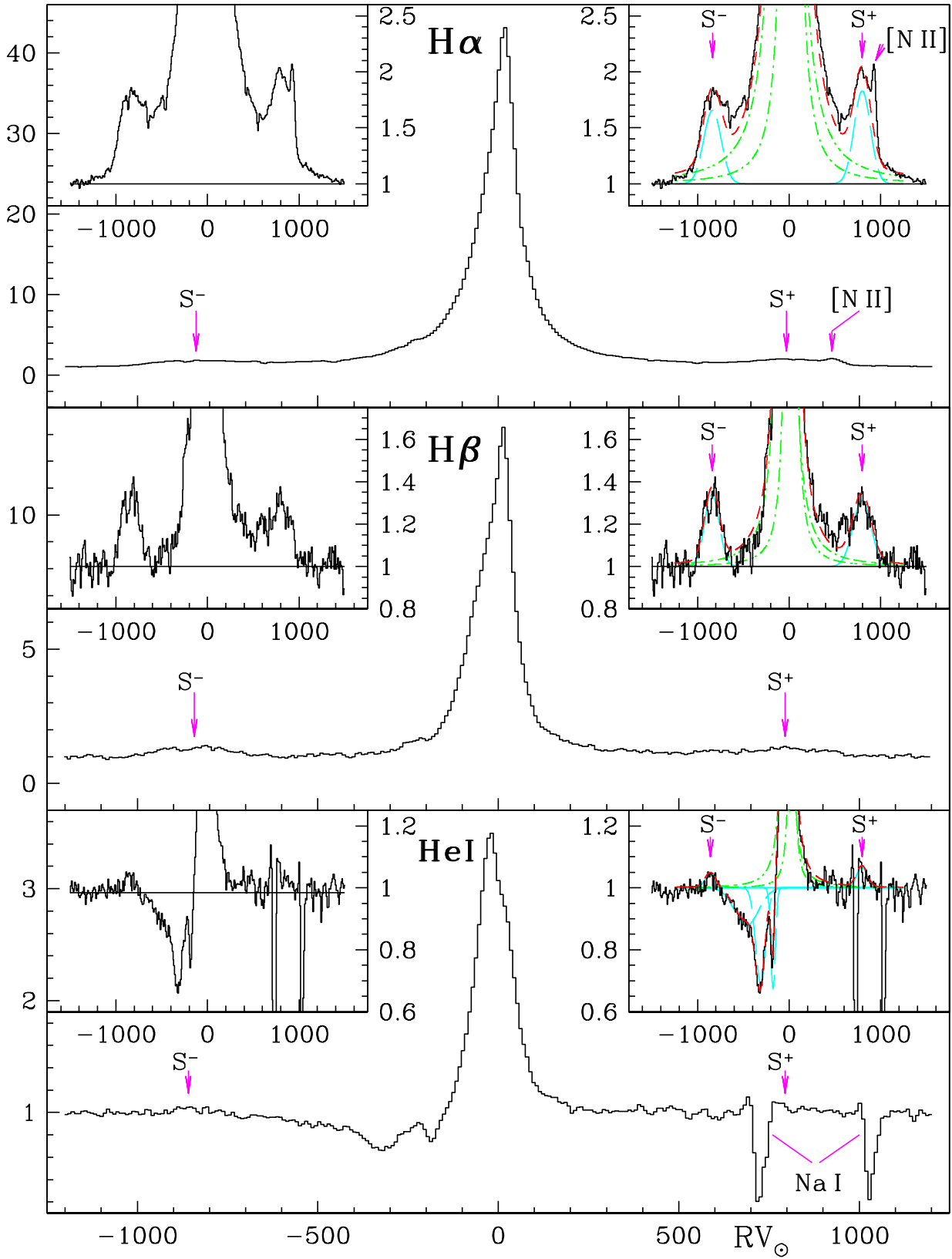

**Fig. 1.** The profiles of the H$\alpha$, H$\beta$ and He I 5876 Å emission lines on the spectrum of Hen 3-1341 for June 8, 1999. The jets emission components are marked with S$^-$ and S$^+$. To emphasize their visibility the profiles are plotted on an expanded ordinate scale on the left side small panels. On the small panels on the right side the fit with two lorentzians (for the main emission lines to account for their asymmetry; dot-dashed lines) and two gaussians (one for each jet component; long dashed lines) is shown. The only mean of such a fit is to allow an accurate measure of the wavelength, FWHM and equivalent width of the jet components as summarized in Table 2.

**Table 1.** Journal of observations.

| Date | HJD 2451000+ | Spectr. | Spectral region |
|---|---|---|---|
| Jun 8, 1999 | 338.463 | 1.82m + Echelle | 4600–9200 Å |
| Aug 18, 1999 | 409.325 | 1.22m + B&C | 6210–6810 Å |
| Aug 22, 1999 | 413.304 | 1.82m + Echelle | 4600–9200 Å |

**Table 2.** Heliocentric radial velocities ($RV_\odot$), equivalent width ($EW_\lambda$) and FWHM of the gaussian fit to the jet emission components for the H$\alpha$, H$\beta$ and He I 5876 Å lines on the June 8, 1999 spectrum of Hen 3-1341.

| | $RV_\odot$ (km s$^{-1}$) | | $EW_\lambda$ (Å) | | FWHM (km s$^{-1}$) | |
|---|---|---|---|---|---|---|
| | $S^-$ | $S^+$ | $S^-$ | $S^+$ | $S^-$ | $S^+$ |
| H$\alpha$ | −837 | +799 | 3.4 | 4.1 | 224 | 212 |
| H$\beta$ | −843 | +795 | 1.1 | 1.3 | 189 | 236 |
| He I | −859 | +795 | 0.13 | 0.20 | 137 | 141 |

Spectrophotometric standard stars have been observed in parallel with Hen 3-1341. The marginally photometric conditions during the observations prevent us from an accurate estimate of the brightness of Hen 3-1341 from our spectra. However, a rough estimate can be derived by the flux level in the spectra and Hen 3-1341 brightness on the screen of the telescope TV guiding system. Both methods suggest that Hen 3-1341 at the time of our observations was $\triangle V \sim 2^{\rm m}\!5$ brighter than the $V = 12^{\rm m}\!94$ reported by Munari et al. (1992) for quiescence.

## 3. Results

Figure 1 presents the profiles of the H$\alpha$, H$\beta$ and He I 5876 Å emission lines for the June 8, 1999 Echelle spectrum. Quite similar profiles result from the August observations listed in Table 1. The outburst state of Hen 3-1341 at the time of our observations is documented in Figures 2 and 3

The Hydrogen Balmer lines show a strong central emission component at a heliocentric radial velocity $RV_\odot$=+16 km s$^{-1}$, with a marked blue asymmetry. The latter is typical of symbiotic stars, with the whole shape being modulated by the orbital motion (Munari 1992). The He I 5876 Å line presents instead a red asymmetry and a two-component P-Cyg profile (similarly to the classical symbiotic star EG And at orbital quadrature, Munari 1992). The P-Cyg signature of mass loss from the outbursting component is visible in the other He I lines too (cf. Figure 3).

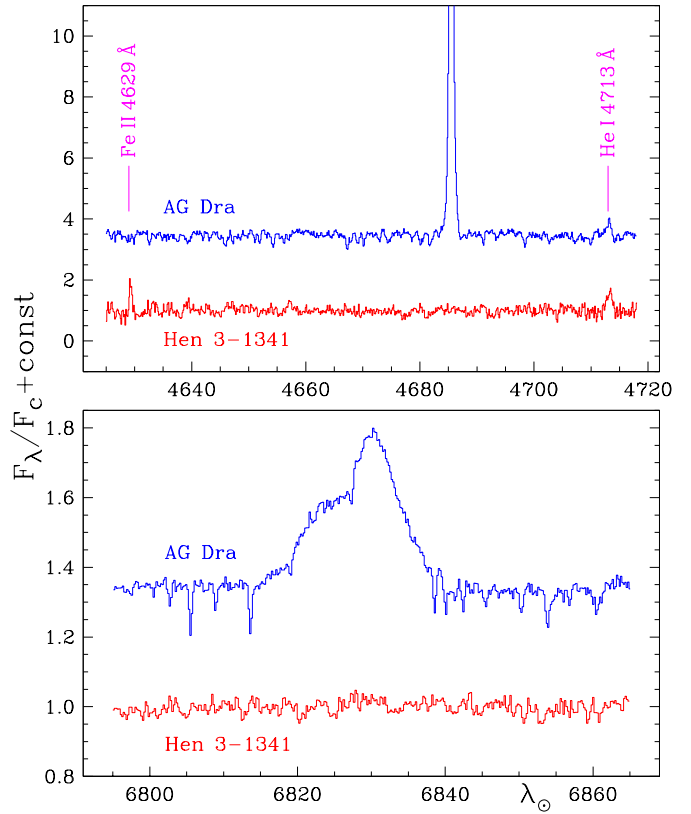

**Fig. 2.** This figures documents the disappearance of high ionization emission lines as a consequence of the outburst state of Hen 3-1341 during our 1999 observations. The spectral regions around He II 4686 Å and the symbiotic band at 6830 Å (see text) for the June 8, 1999 spectrum of Hen 3-1341 are compared to those of the symbiotic star AG Dra observed under the same instrumental conditions on the same night. The absence of the high excitation emission lines is evident.

The central emission lines are flanked on both sides by weaker emission components at $|\Delta RV_\odot| \sim 820$ km s$^{-1}$, with an integrated flux $\sim$3% that of the central emission. We attribute them to bipolar jets. These satellite emissions (hereafter indicated with $S^\pm$) have been fitted with gaussians and the results are plotted in Figure 1 and detailed in Table 2. In spite of the large differences between the main emission lines of Hydrogen and Helium (which complex origin involve the cool giant wind, the outbursting component wind and the circumstellar medium, cf. Munari 1992), the corresponding $S^\pm$ components are much more similar in intensity ratio, radial velocity and shape. This goes with expectations, because the jets are regions highly kinematically confined. The integrated flux of the nearby [N II] 6548 and 6584 Å emission lines (see Figure 1) is $\sim$1% that of $S^-$ and $\sim$3% that of $S^+$, respectively.

According to Livio (1997) the jet velocity in accreting systems is generally of the order of the escape velocity from the central star. If this is assumed to hold true also in the case of symbiotic stars, it would lead to an estimate of the orbital inclination by simple geometry. The orbital inclination could than be used in estimating the opening angle of the jets (see Shahbaz et al. 1997) and therefore to set some contraints on the origin of the jets themselves. Unfortunately, we cannot estimate the escape velocity ($V_{esc} = \sqrt{2\,G\,M_{WD}/R_{WD}}$) mainly because it is unknown if the jets left the white dwarf before or at the onset of the outburst (when $R_{WD} \sim 10^{-2}$ $R_\odot$), or close to outburst maximum (when $R_{WD} \sim 10$ $R_\odot$).

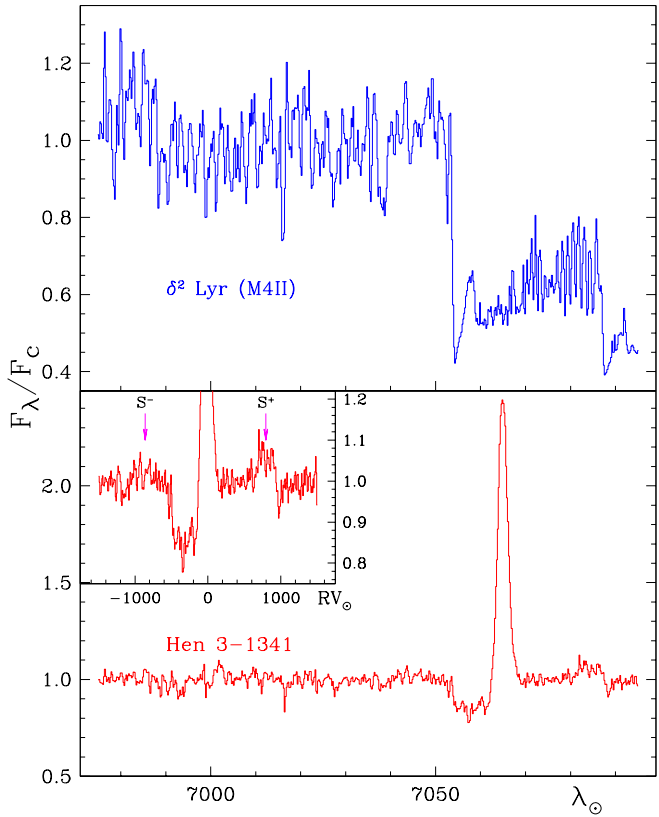

**Fig. 3.** The figure documents the overwhelming of the TiO absorption bands of the M4 giant in Hen 3-1341 by the continuum of the outbursting component (June 8, 1999 spectrum). No sign of the TiO band is left in the outburst spectrum of Hen 3-1341. The P-Cyg absorption and jet emission components in the He I 7065 Å profile are similar to those of He I 5876 Å in Figure 1. The reference M4 II spectrum is taken from the Munari and Tomasella (1992) Echelle atlas of MKK spectral standards.

Among the several follow-up observations that could be carried out in the coming observing season to better characterize the nature and properties of Hen 3-1341 and its jets (if still present), we suggest to consider the following:

(*a*) symbiotic stars with jets show flickering . This is a rare case among normal symbiotic stars (Dobrzycka et al. 1996, and references therein). The properties of the flickering regions are still quite vague (temperature, dimension, location, etc...) and a search in Hen 3-1341 is of considerable interest;

(*b*) high resolution radio/optical/UV imaging of Hen 3-1341 to spatially resolve the jets (as it already happened with CH Cyg and RS Oph) and determine their morphology and follow the evolution;

(*c*) ultraviolet spectroscopy to investigate the mass loss via wind from the white dwarf, its interaction with the slower M4 giant wind and to fix the properties of the white dwarf itself during the outburst (the IUE spectroscopy in quiescence by Gutierrez-Moreno et al. (1997) could be used as a comparison).

*Acknowledgements.* We thank the referee (H.M. Schmid) for the useful comments he has rised on the first version of the paper.

## References

Allen D.A., 1978, MNRAS 184, 601
Allen D.A., 1982, in The Nature of Symbiotic Stars, IAU Coll. 70, M.Friedjung and R.Viotti eds., Reidel, 70, pag. 27
Allen D.A., 1984, Proc.A.S.A. 5, 369
Burgarella D., Paresce F., 1992, ApJ 389, L29
Ciatti F., 1982, in The Nature of Symbiotic Stars, IAU Coll. 70, M.Friedjung and R.Viotti eds., Reidel, 70, pag. 61
Dobrzycka D., Kenyoyn S.J., Milone A.A.E., 1996, AJ 111, 414
Dougherty S.M., Bode M.F., Lloyd H.M., et al., 1995, MNRAS 272, 843
Gutiérrez-Moreno A., Moreno H., Costa E., et al., 1997, ApJ 485, 359
Henize K.G., 1976, ApJS 30, 491
Kenyon S.J., 1986, The symbiotic stars. Cambridge University Press
Livio M., 1997, in Accretion Phenomena and Related Outflows, eds. D.T. Wichramasinghe et al. (San Francisco: ASP Conf. Ser.), p.845
Munari U., 1988, A&A 207, L8
Munari U., 1993, A&A 273, 425
Munari U., Yudin B.F., Taranova O.G., et al., 1992, A&AS 93, 383
Munari U., Tomasella L., 1999, A&AS 137, 521
Mürset U., Schmid H.M., 1999, A&AS 137, 473
Schmid H.M., 1989 A&A 211, L31
Shahbaz T., Livio M., Southwell K.A., 1997, ApJ 484, L59
Shore S., Aufdenberg J.P., Michalitsianos A.G., 1994, AJ 108, 671
Solf J., 1987, A&A 180, 207
Taylor A.R., Seaquist E.R., Mattei J.A., 1986, Nat 319, 38
Taylor A.R., Davis R.J., Porcas R.W., et al., 1989, MNRAS 237, 81
Tomov T., Kolev D., Zamanov R., et al., 1990, Nat 346, 637
Tomov T., Munari U., Kolev D., et al., 1998, A&A 333, L67
Van Winckel H., Duerbeck H.W., Schwarz H.E., 1993, A&AS 102, 401